\documentclass[aps,twocolumn,floatfix,superscriptaddress]{revtex4-1}

\usepackage{bm}
\usepackage{amsmath,amssymb,cases}
\usepackage{braket}
\usepackage{graphicx}
\usepackage{color}

\usepackage{ulem}

\begin{document}

\title{Identification of topological phases using classically-optimized variational quantum eigensolver}

\author{Ken N. Okada}
\thanks{These two authors contributed equally.}
\affiliation{Center for Quantum Information and Quantum Biology (QIQB), Osaka University, Osaka 560-8531, Japan}
\author{Keita Osaki}
\thanks{These two authors contributed equally.}
\affiliation{Graduate School of Engineering Science, Osaka University, Osaka 560-8531, Japan}
\author{Kosuke Mitarai}
\affiliation{Center for Quantum Information and Quantum Biology (QIQB), Osaka University, Osaka 560-8531, Japan}
\affiliation{Graduate School of Engineering Science, Osaka University, Osaka 560-8531, Japan}
\affiliation{JST, PRESTO, Saitama 332-0012, Japan}
\author{Keisuke Fujii}
\affiliation{Center for Quantum Information and Quantum Biology (QIQB), Osaka University, Osaka 560-8531, Japan}
\affiliation{Graduate School of Engineering Science, Osaka University, Osaka 560-8531, Japan}
\affiliation{RIKEN Center for Quantum Computing (RQC), Saitama 351-0198, Japan}

\begin{abstract}
Variational quantum eigensolver (VQE) is regarded as a promising candidate of hybrid quantum-classical algorithm for the near-term quantum computers.
Meanwhile, VQE is confronted with a challenge that statistical error associated with the measurement as well as systematic error could significantly hamper the optimization. 
To circumvent this issue, we propose classically-optimized VQE (co-VQE), where the whole process of the optimization is efficiently conducted on a classical computer. 
The efficacy of the method is guaranteed by the observation that quantum circuits with a constant (or logarithmic) depth are classically tractable via simulations of local subsystems. 
In co-VQE, we only use quantum computers to measure nonlocal quantities after the parameters are optimized.
As proof-of-concepts, we present numerical experiments on quantum spin models with topological phases. 
After the optimization, we identify the topological phases by nonlocal order parameters as well as unsupervised machine learning on inner products between quantum states.
The proposed method maximizes the advantage of using quantum computers while avoiding strenuous optimization on noisy quantum devices.
Furthermore, in terms of quantum machine learning, our study shows an intriguing approach that employs quantum computers to generate data of quantum systems while using classical computers for the learning process.
\end{abstract}

\maketitle

\section{Introduction}\label{sec:intro}
Over the past few years, quantum computation has garnered a lot of interest from various scientific fields because of rapid technological progress in quantum device manufacturing \cite{Kjaergaard2020, Bruzewicz2019}. 
To date, quantum computers with several tens of qubits have been experimentally realized \cite{Arute2019}, which leads to an expectation that a few hundreds of qubits could be implemented within the next decade.
Quantum computers with that scale are named noisy intermediate-scale quantum (NISQ) devices \cite{Preskill2018} in a sense that they have a classically-intractable number of qubits, but that they are vulnerable to noise due to lack of quantum error correction.
As for software development, researchers have been focused on how to utilize NISQ devices to solve problems that are hard to tackle on classical computers \cite{McArdle2020, Endo2021, Cerezo2021, bharti2021}. 
They have developed various hybrid quantum-classical algorithms applicable to quantum many-body problems \cite{Peruzzo2014, McClean2016}, combinatorial optimization \cite{Farhi2014}, machine learning \cite{Mitarai2018, Farhi2018, Havlicek2019, Kusumoto2021, Schuld2019, Benedetti2019}, and so on. 

Variational quantum eigensolver (VQE) \cite{Peruzzo2014, McClean2016} is one of the most promising hybrid quantum-classical algorithms to solve quantum many-body problems.
VQE, based on the variational principle, searches for the ground state of the target Hamiltonian with a parameterized quantum circuit used as the ansatz.
In VQE, one measures energy or its parameter derivatives on a quantum computer and accordingly updates variational parameters on a classical computer.
One repeats this process until the energy converges to a minimum.
So far, benchmark experiments using actual quantum devices have demonstrated that with appropriate ans\"{a}tze, VQE could yield approximate solutions with high precision for small quantum systems, primarily, small molecules \cite{Peruzzo2014, OMalley2016, Kandala2017, Colless2018, Kandala2019, Shen2017, Hempel2018, Kokail2019, Nam2020}. 
Meanwhile, they have also revealed that optimization process in VQE could be significantly affected by statistical error as well as systematic error; the former intrinsically arises from measurement on quantum circuit, whereas the latter comes from imperfect fidelity of quantum gates and read-out or decoherence of quantum states.
Although various techniques of error mitigation have been proposed \cite{Li2017, Temme2017, Endo2018, McClean2017, Bonet-Monroig2018, McArdle2019, Chen2019, Maciejewski2020, Kwon2021, Strikis2020, Czarnik2020, Endo2021}, some of which have been experimentally found effective to a certain extent \cite{Kandala2019, Colless2018}, it still remains an issue how to minimize adverse effect of error on computation with NISQ devices. 

We propose a variant of VQE that conducts the whole process of optimization on a classical computer. 
We term this method classically-optimized VQE (co-VQE). 
By its nature, co-VQE is free of the aforementioned errors because it does not involve a quantum computer in the optimization process.
It is based on an observation that shallow depth circuits are easy to simulate when we are only interested in local observables.
For Hamiltonians with only local interactions, this allows us to optimize the ansatz classically without exponential computational cost.
Another important observation is that, while local observables on shallow circuits are classically simulatable, nonlocal ones are not in general.
Therefore, we need to run the optimized circuit on a quantum device if we wish to obtain information about nonlocal quantities.
This kind of situation occurs very naturally with a quantum system which exhibits a topological phase transition, where the system Hamiltonians are local but its order parameters are global  \cite{Wen1990, Gu2009, Pollmann2012}.
Another useful quantity that has to be measured on a quantum device is inner product between the states, which can be used for, e.g., the unsupervised clustering of states in different phases \cite{Yang2021}. 

As numerical demonstrations of co-VQE, we study two quantum spin models in which topological orders emerge; the one-dimensional (1D) cluster model \cite{Doherty2009, Cong2019} and the two-dimensional (2D) toric code model with transverse field \cite{Trebst2007, Blote2002, Wu2012}.
co-VQE would be practically important in 2D systems because in 1D systems, one can often obtain the ground states with high accuracy as well as evaluate nonlocal observables efficiently on classical computers using matrix product states (MPS).
After optimization, we characterize the topological phases by measuring nonlocal order parameters. 
Moreover, for the 1D model, we identify the topological phase via another method, measurement of fidelity combined with an unsupervised machine learning.
Here quantities of quantum systems are created by quantum computers and processed by classical computers.
This could be another promising type of quantum machine learning in the NISQ era.
As compared to other near-term quantum machine learning algorithms \cite{Mitarai2018, Farhi2018, Havlicek2019, Kusumoto2021, Schuld2019, Benedetti2019} treating classical data, we might say that our proposal makes natural use of quantum computers.
We also refer to recent similar works in Refs. \cite{Sancho-Lorente2021, Wu2021}. 

The paper is organized as follows.
First, we describe the principle of co-VQE in Sec. \ref{sec:vqe_co}.
We devote Secs. \ref{sec:cm} and \ref{sec:tcm} to numerical demonstrations of co-VQE on the 1D cluster and 2D toric code models with transverse field, respectively.
Finally, we summarize our results in Sec. \ref{sec:conclusion}.

\section{Classically-optimized variational quantum eigensolver}\label{sec:vqe_co}
\begin{figure}
\begin{center}
\includegraphics[width=\columnwidth]{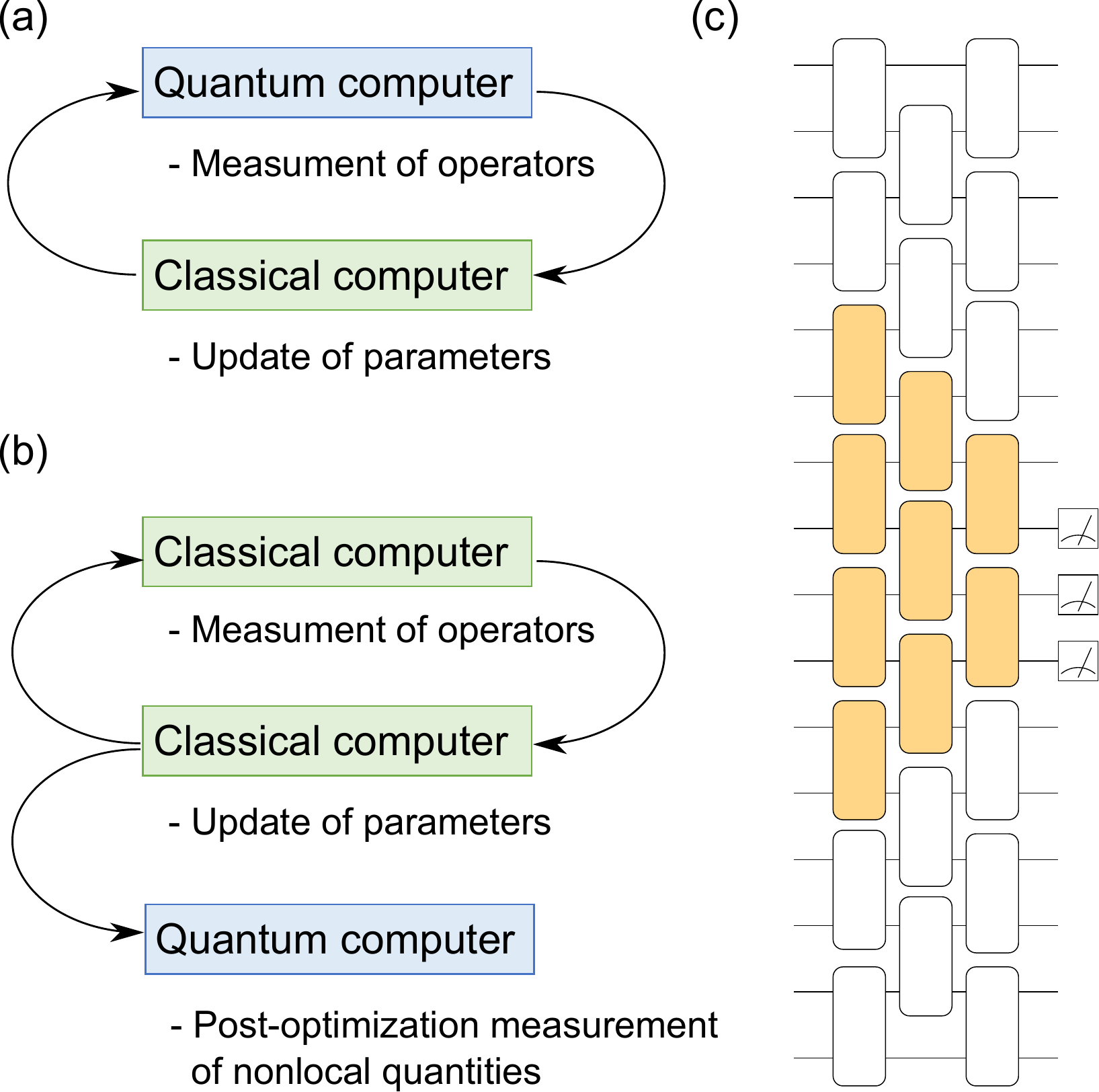}
\end{center}
\caption{Schematic diagram of (a) the original VQE and (b) co-VQE. (c) Causal cone (orange area)  in a brickwall quantum circuit.}
\label{fig:schematic}
\end{figure}

In this section, we explain the premise and principle of co-VQE.
In the optimization process of the original VQE, one evaluates expectation values of operators by measurement on a quantum computer and accordingly update variational parameters on a classical computer [Fig. \ref{fig:schematic}(a)]. 
In co-VQE, meanwhile, one efficiently calculates the expectation values with a classical computer instead of using a quantum computer [Fig. \ref{fig:schematic}(b)].
The key to the efficiency is reduction in the number of simulated qubits, thanks to the locality of causal cones in shallow-depth quantum circuits [Fig. \ref{fig:schematic}(c)].

As prerequisites for co-VQE, we suppose the following constraints on the target problem and ansatz.
\begin{enumerate}
 \item Operators measured in the optimization process are local. \label{operator_locality}
 \item The initial state of the quantum circuit is a product state or more generally a stabilizer state. \label{init_product}
 \item The quantum circuit is only composed of local operations. \label{circuit_locality}
 \item The quantum circuit has a constant (or logarithmic) depth as a function of the system size. \label{constant_depth}
\end{enumerate}
In many systems of interest in condensed-matter physics, the constraint (i) holds for operators composing the physical Hamiltonian. 
In addition, a majority of the circuit ans\"{a}tze proposed so far meet the constraints (ii) and (iii).
The constraint (iv) allows for hardware-efficient ans\"{a}tze \cite{Peruzzo2014, Kandala2017, Kandala2019} as well as the Hamiltonian variational ansatz \cite{Wecker2015} in spin models as we treat in the paper.
Although (iv) restricts depth of the ansatz, we remark that, when one increases the system size up to, e.g., $10^3$ or $10^4$, in the NISQ era, polynomially-scaled circuit depths are not ideal or viable due to exponential decrease in fidelity. 
We also point out that quantum circuits with constant depth can have quantum advantages \cite{Bravyi2018}.

Under those constraints, one can classically simulate quantum circuits without suffering from exponential increase in the computational cost.
As a simple example, let us consider a 1D brickwall quantum circuit applied to a product state [Fig. \ref{fig:schematic}(c)].
As indicated in Fig. \ref{fig:schematic}(c), measurement of a local operator on the circuit only involves a part of the qubits considering the causal cone.
Therefore, it only suffices to simulate a local subsystem rather than the entire system.
The argument can be generalized to $d$ dimensions ($d\geq 1$).
Let $N$ be the number of qubits of the entire system.
Suppose we calculate the expectation value of a local observable $\mathcal{O}$ with respect to a quantum circuit with depth $l$.
Assuming that $\mathcal{O}$ is described as a sum of $m$ $k$-body Pauli operators, the maximal number of qubits among the causal cones corresponding to $m$ Pauli operators, $M$, is estimated as $M=O(kl^d)$.
Hence, evaluation of the observable $\mathcal{O}$ costs $O(m2^{kl^d})$, whereas it costs $O(m2^N)$ when one simulates the entire system.
In the case where $\mathcal{O}$ is the Hamiltonian, considering $m=O({\rm poly}(N))$ ($O(N)$ in typical spin models), the simulation assisted by causal cones can be performed efficiently as long as the depth $l$ is constant (or logarithmic) with respect to $N$.
Importantly, this technique enables us to classically simulate systems with more qubits than current computers could afford to simulate, e.g., $N\sim 100$ as long as $M$ is sufficiently small. 

One might think that the efficient simulatability of low-depth quantum circuits for local observables is limited to cases where the initial state is a product state.
In fact, however, it holds as long as the initial state is a stabilizer state.
Supposing that $U$ represents the quantum circuit, $U^{\dag}\mathcal{O}U$ can be decomposed into $O(m4^{kl^d})$ $(kl^d)$-body Pauli operators.
Since expectation values of Pauli products with respect to a stabilizer state can be efficiently calculated on classical computers, one can efficiently simulate the expectation value of the local observable when $l$ is constant (or logarithmic).
In the numerical simulations in Secs. \ref{sec:cm} and \ref{sec:tcm}, we use stabilizer states as the initial state of the ansatz.

The argument above raises a question; once that the entire optimization is conducted on a classical computer, in what cases does one benefit from using a quantum computer?
The answer would be in cases where one conducts measurement of nonlocal quantities [Fig. \ref{fig:schematic}(b)].
One case which necessitates such measurements would be characterization of topological orders \cite{Wen1990, Gu2009, Pollmann2012} emergent in certain quantum spin models.
Since order parameters of topological phases are nonlocal, it is necessary to measure them on a quantum device to explicitly tract topological orders.
Another case that we consider here is measurement of fidelity, the absolute value of inner product between two quantum states.
Fidelity is useful to detect quantum phase transitions, especially when one does not know an apparent form of order parameter \cite{Quan2006, Zanardi2006}.
We also need to evaluate it on a quantum device due to its globality.

In Secs. \ref{sec:cm} and \ref{sec:tcm}, we present proof-of-concept demonstrations of co-VQE on quantum spin models with topological phases.
First, in Sec. \ref{sec:cm}, we study the 1D cluster model with transverse field.
However, in 1D systems, one can efficiently evaluate nonlocal observables on a classical computer if the circuit has a constant or logarithmic depth.
In 2D systems, on the other hand, it is not feasible to classically simulate nonlocal observables even with a constant depth.
In that sense, co-VQE is expected to provide practical importance in 2D systems.
With that in mind, in Sec. \ref{sec:tcm}, we treat the 2D toric code model with transverse field.

We simulate quantum circuits using Qulacs \cite{Suzuki2020} and optimize variational parameters using the BFGS method implemeted in the SciPy library \cite{Virtanen2020}.
We also perform exact diagonalization (ED) to obtain reference values.
For clustering, we employ scikit-learn \cite{Pedregosa2011}.
We note that our studies are restricted to a classically-tractable number of qubits ($N\leq25$) so that we could simulate nonlocal measurements, which would be conducted on quantum computers in actual usage of the co-VQE.

\section{Transverse-field cluster model}\label{sec:cm}

\subsection{Nonlocal order parameter}\label{sec:cm_order_para}
In this section, we study the 1D cluster model with transverse field \cite{Doherty2009}. 
We consider an $N$-qubit chain with open boundary condition, where the Hamiltonian reads
\begin{equation}
\mathcal{H}_{\rm cluster} = -\sum_{i=2}^{N-1} K_i - J\sum_{i=1}^N X_i.
\label{eq:cm_hamiltonian}
\end{equation}
We define the stabilizer $K_i$ as $K_i=Z_{i-1}X_iZ_{i+1}$.
For $J=0$, the ground state of the model is in an symmetry-protected topological (SPT) phase   \cite{Gu2009, Pollmann2012} and called the cluster state \cite{Raussendorf2001}. 
The state is characterized with eigenvalues of $K_i$ being equal to $1$ for all $i$.
The model is exactly solvable because it can be mapped to the transverse-field Ising model via a nonlocal transformation \cite{Doherty2009}. 
The exact solution tells us that as $J$ increases from zero, the cluster state transitions to a trivial phase at $J=1$. 
The order parameter characterizing the phase transition is a product of the stabilizers represented as
\begin{equation}
\Omega=\Braket{\prod_{k=1}^{\left[\frac{N-1}{2}\right]}K_{2k}}.
\label{eq:CM_order_para}
\end{equation} 
The cluster state corresponds to $\Omega=1$, whereas the trivial phase has $\Omega=0$ \cite{Doherty2009}.
We remark that the order parameter $\Omega$ is a nonlocal observable.

\begin{figure}
\begin{center}
\includegraphics[width=\columnwidth]{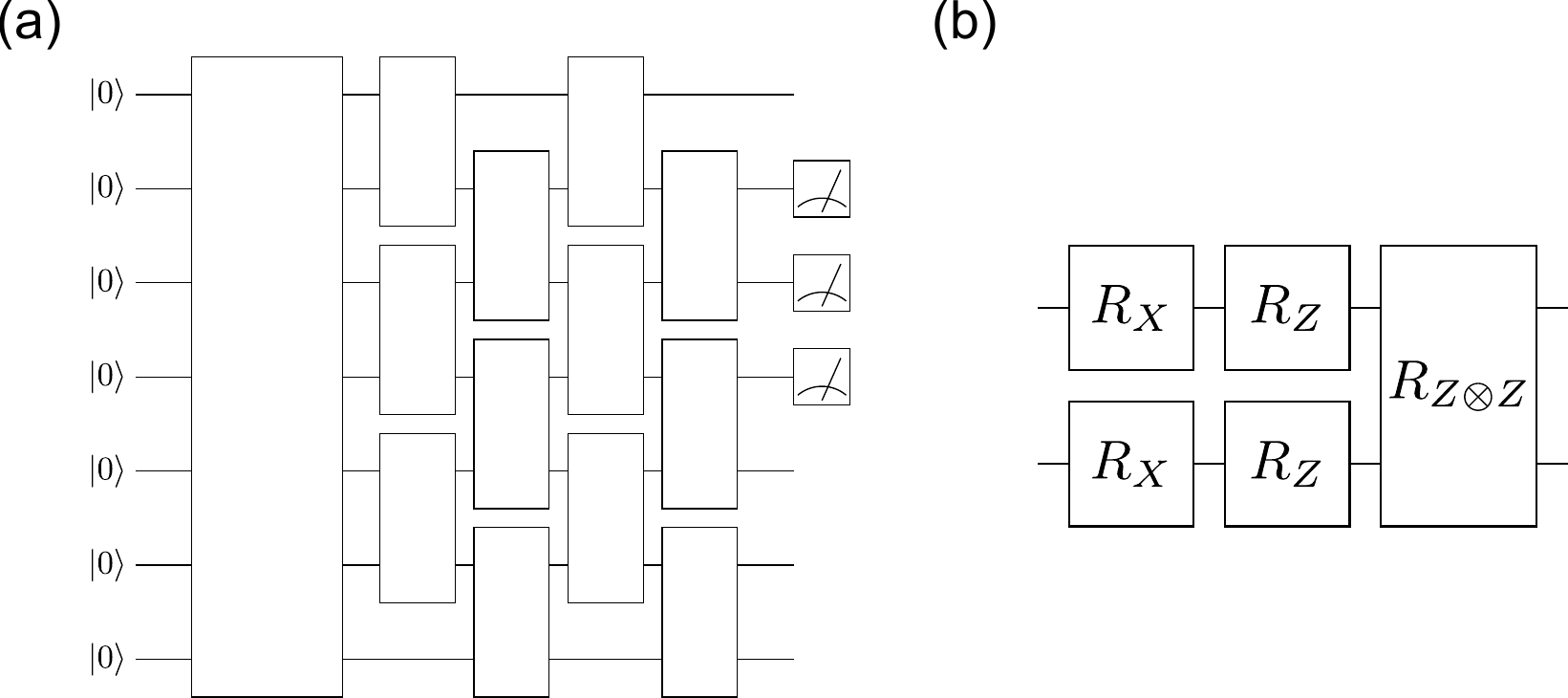}
\end{center}
\caption{(a) Brickwall quantum circuit used as the ansatz of co-VQE for the transverse-field cluster model. The leftmost block corresponds to a circuit that generates the cluster state. (b) Quantum operations composing a single brick in (a).}
\label{fig:CM_circuit}
\end{figure}

We study the ground state of $\mathcal{H}_{\rm cluster}$ by co-VQE.
We use a brickwall quantum circuit applied to the cluster state [Fig. \ref{fig:CM_circuit}(a)] as the ansatz. 
The ansatz and Hamiltonian meet four conditions for co-VQE listed in Sec. \ref{sec:vqe_co}. Note that the cluster state is a stabilizer state.
As shown in Fig. \ref{fig:CM_circuit}(b), each brick is composed of four single-qubit and one two-qubit rotations with independent variational parameters. 
We define the depth of the circuit $d$ as the number of brick layers ($d=4$ in Fig. \ref{fig:CM_circuit}(a)).
Importantly, the ansatz corresponds to the Hamiltonian variational ansatz \cite{Wecker2015}.
This guarantees that the ansatz can generate the exact solution when $d\to\infty$. 
As described in Sec. \ref{sec:vqe_co}, in the optimization process of co-VQE, we only simulate relevant qubits dictated by the causal cone [Fig. \ref{fig:schematic}(c)] to evaluate expectation values of the local terms in $\mathcal{H}_{\rm cluster}$. 

\begin{figure}[htb]
\begin{center}
\includegraphics[width=0.8\columnwidth]{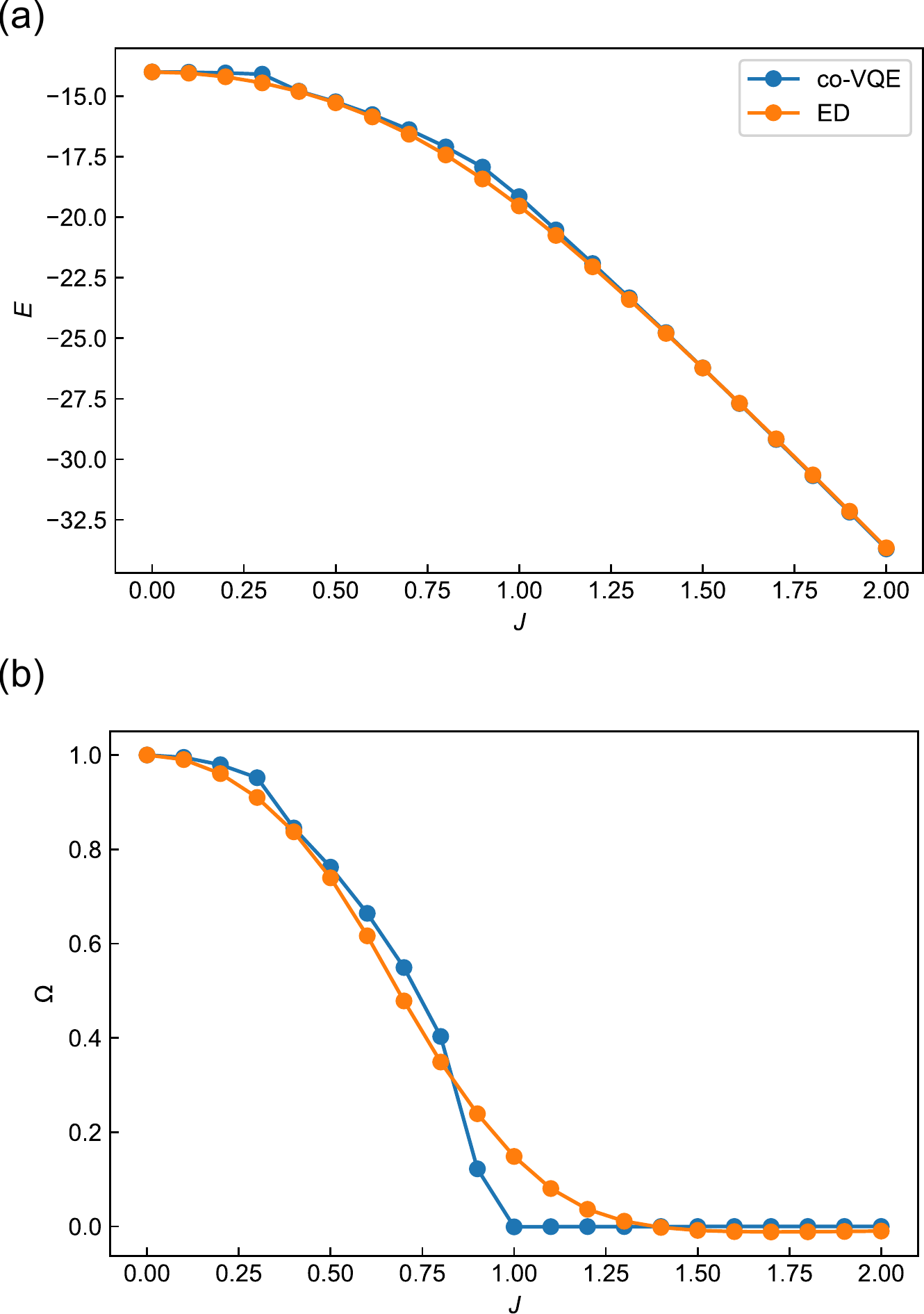}
\end{center}
\caption{$J$ dependence of (a) the energy $E$ and (b) nonlocal order parameter $\Omega$ in Eq. (\ref{eq:CM_order_para}) of the ground states obtained by co-VQE on the transverse-field cluster model with the Hamiltonian $\mathcal{H}_{\rm cluster}$ in Eq. (\ref{eq:cm_hamiltonian}). The blue (orange) points show the data of co-VQE (ED). In co-VQE, we set the depth of the ansatz as $d=4$. The calculations are done for $N=16$.}
\label{fig:CM_energy}
\end{figure}

First, we study the phase transition by explicitly measuring the nonlocal order parameter $\Omega$.
For each value of $J$ ($0\leq J\leq2$), we optimize variational parameters and then compute $\Omega$.
In Figs. \ref{fig:CM_energy}(a) and \ref{fig:CM_energy}(b), we show $J$ dependence of the ground state energy $E$ and $\Omega$, respectively, for $N=16$ and $d=4$ as well as those calculated by ED.
As expected, the order parameter $\Omega$ of ED [Fig. \ref{fig:CM_energy}(b)] points to the phase transition at $J\sim 1$.
Figures \ref{fig:CM_energy}(a) and \ref{fig:CM_energy}(b) show that both the energy and order parameter of co-VQE agree well with those of ED.
We note that $\Omega$ of co-VQE shows a slightly steeper decline at the phase transition than that of ED [Fig. \ref{fig:CM_energy}(b)], which might be attributed to limitation of expressibility in the low-depth ansatz. 
The result implies that the brickwall circuit of even a low depth [Fig. \ref{fig:CM_circuit}(a)] can capture a broad range of quantum states including the cluster state and trivial state at the end points.

\subsection{Clustering on fidelities}\label{sec:cm_fidelity}

\begin{figure}[htb]
\begin{center}
\includegraphics[width=\columnwidth]{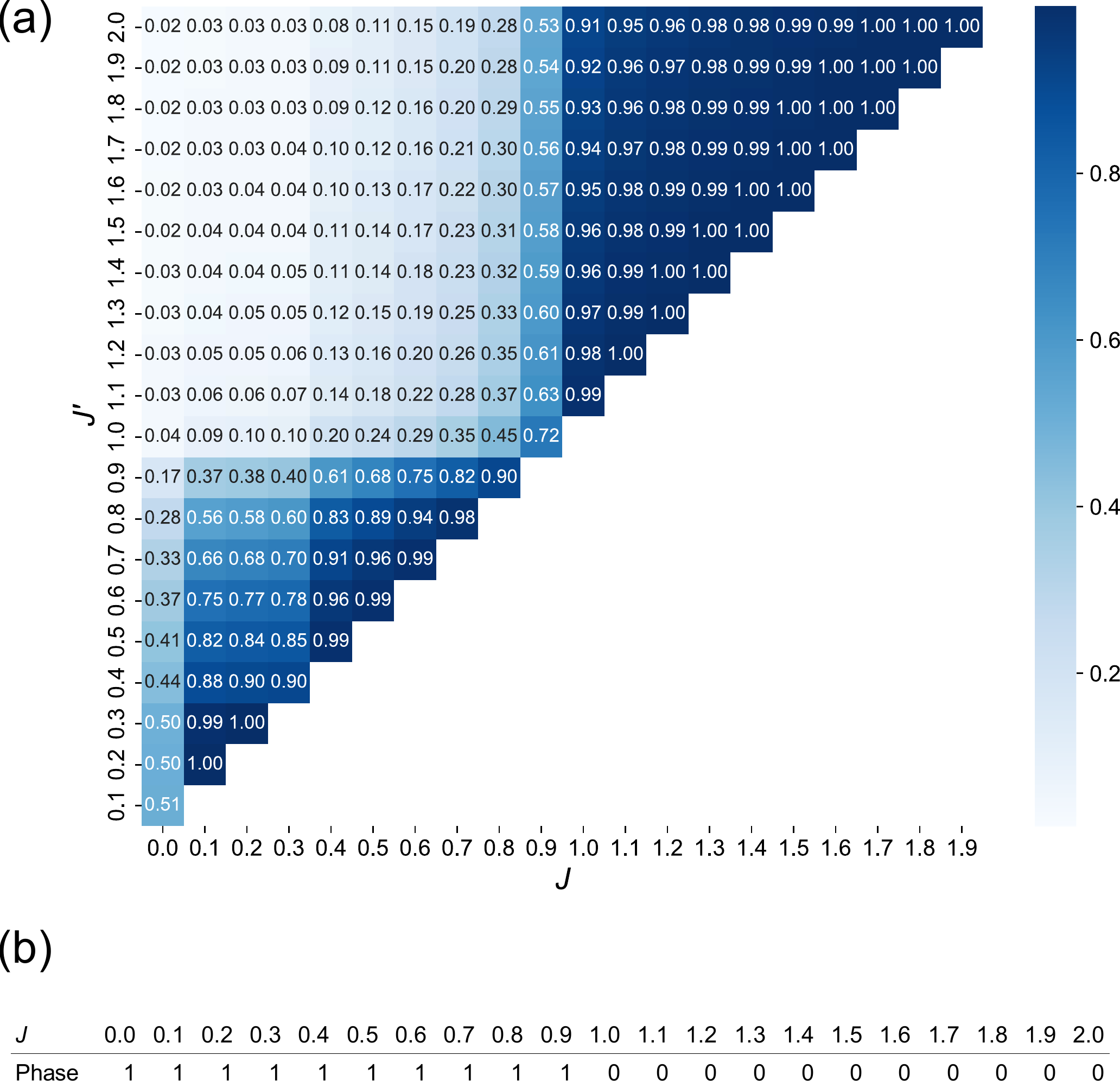}
\end{center}
\caption{(a) Color plot of the fidelity between the states optimized by co-VQE for different values of $J$. (b) Phase classification by spectral clustering on the data set in (a). Phase $1$ corresponds to the SPT phase and $0$ to the trivial phase.}
\label{fig:fidelity}
\end{figure}

Next, as another approach to detect the phase transition, we use fidelity for unsupervised machine learning.
Fidelity measures similarity between two quantum states $\ket{\Psi}$ and $\ket{\Phi}$, and it is defined as
\begin{equation}
F=\left|\braket{\Phi|\Psi}\right|.
\label{eq:fidelity}
\end{equation} 
In the last decade and a half, researchers in condensed-matter physics have found out that fidelity is a useful tool to detect quantum phase transitions \cite{Quan2006, Zanardi2006}, especially when one does not have information of order parameters.
This stems from the property that the fidelity $F$ equals almost one between quantum states within the same phase, whereas zero between states in different phases.
Although in the thermodynamic limit, one should employ other relevant quantities such as fidelity per site \cite{Zhou2008a, Zhou2008b, Sancho-Lorente2021} or fidelity susceptibility \cite{You2007} to avoid the orthogonality catastrophe \cite{Anderson1967}, in the following we can safely rely on fidelity because we only treat small-size systems.

We compute fidelities between the optimized states for different values of $J$.
Figure \ref{fig:fidelity}(a) shows the color plot of fidelity $\left|\braket{\Psi(J^\prime)|\Psi(J)}\right|$.
The top row of the color plot in Fig. \ref{fig:fidelity}(a) shows that as $J$ increases, the fidelity $\left|\braket{\Psi(J^\prime=2.0)|\Psi(J)}\right|$ steeply arises around $J=1$.   
Therefore one can speculate that a phase transition seems to take place around $J=1$.

To confirm the observation, we conduct clustering on the data set of fidelity for phase classification [Fig. \ref{fig:fidelity}(a)].
Here, we apply the spectral clustering method with the number of clusters fixed at two while using the fidelity as affinity between data points.
As a consequence, we obtain two phases as shown in Fig. \ref{fig:fidelity}(b); one corresponds to $0\leq J\leq0.9$ (SPT phase), and the other to $1\leq J\leq2$ (trivial phase).
The result is consistent with the exact phase diagram.
We emphasize that clustering allows us to detect the phase transition without prior knowledge of the order parameter $\Omega$.
Our clustering analysis contrasts with recent studies that have proposed similar methods for classifying phases based on measurement of fidelity in conjunction with classical machine learning \cite{Sancho-Lorente2021, Wu2021}, both of which employ supervised machine learning.

\section{Transverse-field toric code model}\label{sec:tcm}
\begin{figure}[htb]
\begin{center}
\includegraphics[width=0.5\columnwidth]{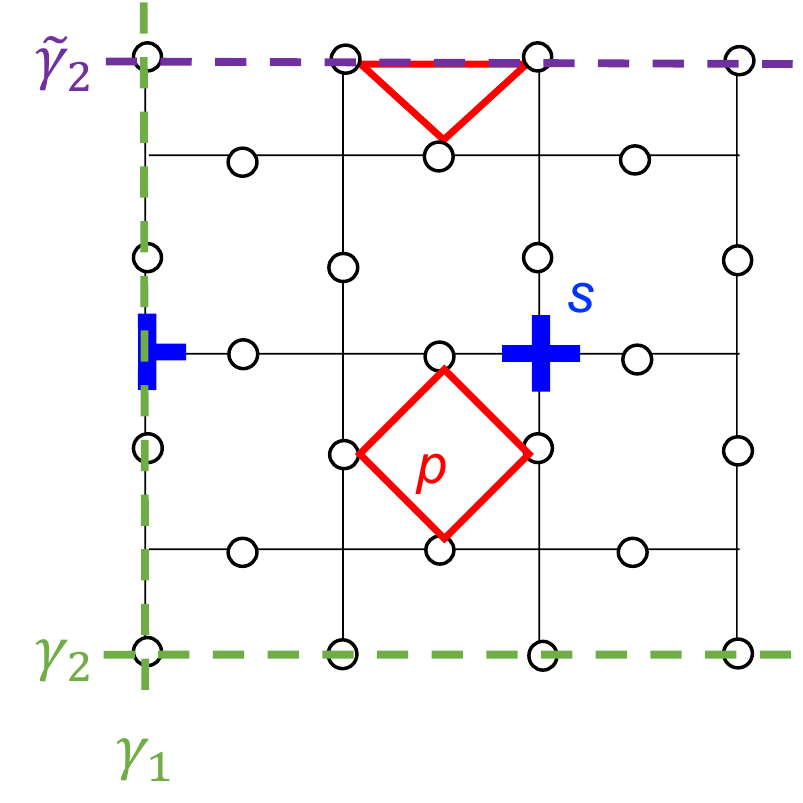}
\end{center}
\caption{Square array of qubits (open circle) with $L=4$ for the toric code model.}
\label{fig:square}
\end{figure}
While we have considered the 1D model in the previous section, classical approaches using MPS or multiscale entanglement renormalization ansatz (MERA) are in many cases sufficient for 1D models.
In this section, we study the toric code model with transverse field as a nontrivial example in two dimensions.
The model is defined on a square array of $N=L^2+(L-1)^2$ ($L$: linear dimension) qubits with open boundary conditions as shown in Fig. \ref{fig:square}.
The Hamiltonian reads
\begin{equation}
\mathcal{H}_{\rm toric}=-\sum_sA_s-\sum_pB_p-h_z\sum_{i=1}^NZ_i,
\label{eq:tcm_hamiltonian}
\end{equation}
where $A_s$ and $B_p$ are stabilizers defined as $A_s=\prod_{i\in s}X_i$ and $B_p=\prod_{i\in p}Z_i$ for each square $s$ and plaquette $p$ [Fig. \ref{fig:square}], and $h_z$ represents strength of transverse field.
For $h_z=0$, the ground states are toric code states with a nontrivial topology \cite{Wen1990}. They are characterized with eigenvalues of $A_s$ and $B_p$ being equal to $1$ for all $s$ and $p$, which are two-fold degenerate in our setting of boundary conditions.
These two-fold ground states are distinguished by eigenvalues of a logical operator $L_Z$, defined as $L_Z=\prod_{i\in\gamma_1}Z_i$ [Fig. \ref{fig:square}].
With application of $h_z$, quantum Monte Carlo calculations revealed that the toric code state undergoes a topological transition at $h_z=0.328474(3)$ in the thermodynamic limit \cite{Trebst2007, Blote2002, Wu2012}.

Below we study the model by co-VQE.
We use the Hamiltonian variational ansatz \cite{Wecker2015} in a more explicit form than in Sec. \ref{sec:cm}.
The ansatz is expressed with variational parameters $\beta_l$ and $\gamma_l$ as
\begin{equation}
\ket{\Psi}=\prod_{l=1}^{D}\left[e^{-i\beta_l\mathcal{H}^0_{\rm toric}}e^{-i\gamma_l\mathcal{H}^1_{\rm toric}}\right]\ket{\Psi(L_Z=1)},
\label{eq:tcm_ansatz}
\end{equation}
where both $\mathcal{H}^0_{\rm toric}$ and  $\mathcal{H}^1_{\rm toric}$ are parts of $\mathcal{H}_{\rm toric}$ defined as $\mathcal{H}^0_{\rm toric}=-\sum_sA_s-\sum_pB_p$ and $\mathcal{H}^1_{\rm toric}=-h_z\sum_{i=1}^NZ_i$.
The initial state $\ket{\Psi(L_Z=1)}$ represents the toric code state with the eiganvalue of $L_Z$ being equal to $1$, which is the exact ground state of $\mathcal{H}_{\rm toric}$ with $h_z=0$.
The Hamiltonian and ansatz satisfy four constraints for co-VQE listed in Sec. \ref{sec:vqe_co}.

The reason why we set the toric code state as the initial state is because topologically ordered states cannot be generated from a product state with a constant depth circuit. 
By doing so, one can study the region belonging to the same topological phase with a shallow quantum circuit.
A preparation of the toric code state $\ket{\Psi(L_Z=1)}$ has been discussed in the context of quantum error correction. More precisely, the syndrome measurements for $A_s$ project a product state $\ket{0}$ onto the toric code state $\ket{\Psi(L_Z=1)}$. The randomness of the measurement outcomes can be treated classically by updating the Pauli frame and rewriting the state and operations in an appropriate basis.

\begin{figure}[htb]
\begin{center}
\includegraphics[width=0.8\columnwidth]{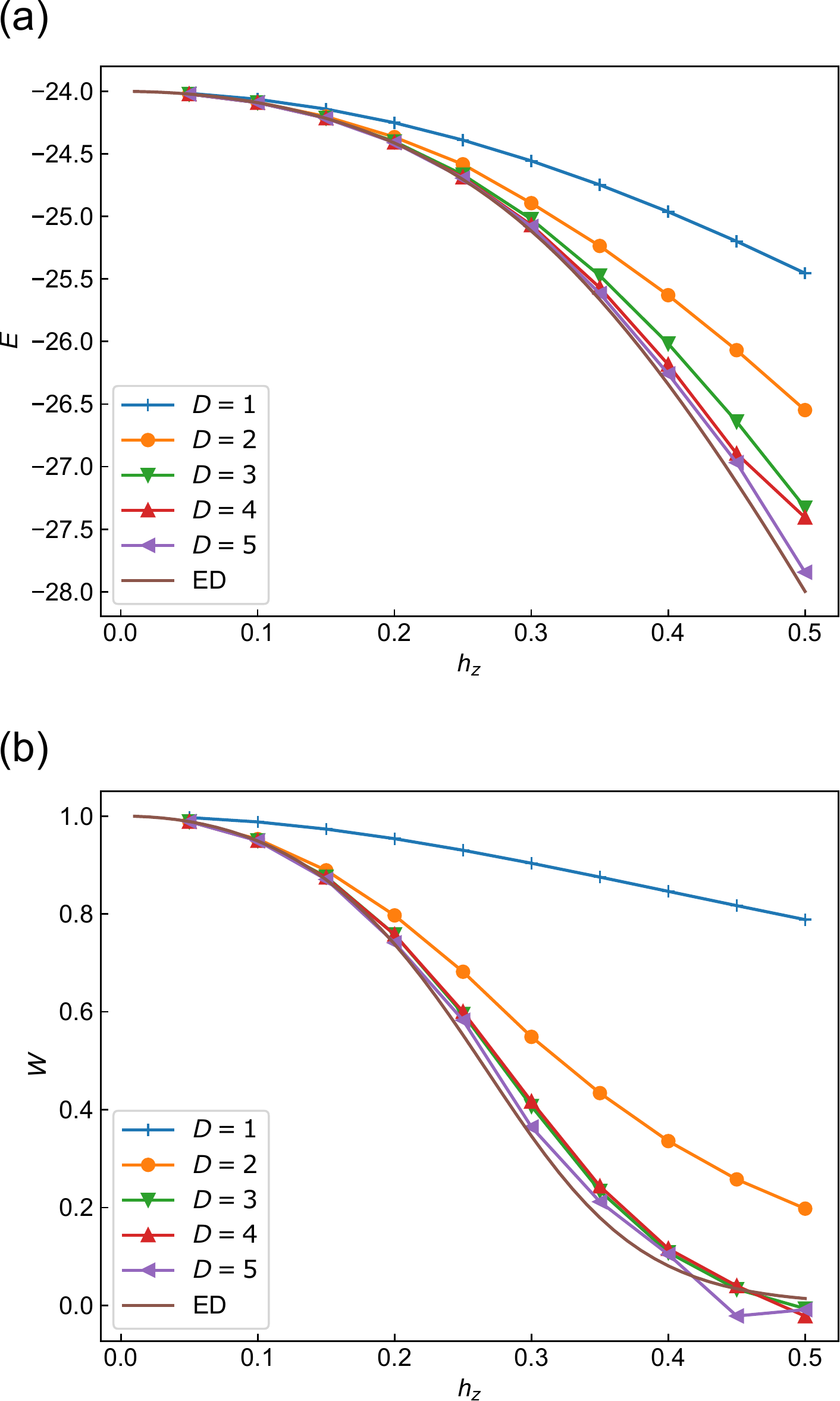}
\end{center}
\caption{$h_z$ dependence of (a) the energy and (b) nonlocal order parameter $W$ in Eq. (\ref{eq:tcm_order_para}) of the ground states obtained by co-VQE on the transverse-field toric code model with the Hamiltonian $\mathcal{H}_{\rm toric}$ in Eq. (\ref{eq:tcm_hamiltonian}). The purple line shows the data of ED. The calculations are done for $L=4$ ($N=25$).}
\label{fig:tcm_energy}
\end{figure}

To tract the topological order, we evaluate the Wilson loop operator $W$ \cite{Yu2008}, which we define as [Fig. \ref{fig:square}]
\begin{equation}
W=\prod_sA_s=\prod_{i\in\gamma_2,\tilde{\gamma}_2}X_i.
\label{eq:tcm_order_para}
\end{equation}
In Figs. \ref{fig:tcm_energy}(a) and (b), we present $h_z$ dependence of the energy $E$ and nonlocal order parameter $W$, respectively, calculated by co-VQE together with those of ED. 
The calculations are done for $L=4$ ($N=25$).
The loop operator $W$ computed by ED [Fig. \ref{fig:tcm_energy}(b)] shows that the toric code state gradually transitions to a trivial state as $h_z$ increases up to 0.5.
Figure \ref{fig:tcm_energy}(a) shows that the energy of co-VQE becomes lower as the depth $D$ increases and nearly overlaps with that of ED for $D=5$.
With regard to $W$, the result of co-VQE almost replicates that of ED with a lower depth ($D=3$) [Fig. \ref{fig:tcm_energy}(b)].

Our calculations indicate that the ansatz with the toric code state as the initial state is effective.
Specifically, for small external field, it may describe the ground states with a constant depth.
Although we restricted ourselves into a simulatable number of qubits in the calculations above, co-VQE would be beneficial for larger $N$ by using real quantum devices.
When you evaluate the energy, the maximal number of qubits among all the causal cones is represented as $M=4(d+1)^2$.
Therefore, classical optimization would be possible up to $d=2$ irrespective of $N$.
If one intends to increase the depth, i.e.,  $d\geq3$, one needs to conduct optimization on quantum computers, but even in that case, co-VQE for $d=2$ would help one to determine the initial parameters for $d=3$ and thus reduce the risk of being trapped at suboptimal solutions.

\section{Conclusion}\label{sec:conclusion}
In this work, we propose co-VQE, a variant of VQE that is more reliant on classical computers. 
In co-VQE, assuming locality of the Hamiltonian and constant (or logarithmic) depth of the ansatz, the whole process of optimization is efficiently conducted on a classical computer.
The efficiency of the classical optimization comes from exponential reduction in simulation costs by virtue of causal cones in quantum circuits.
Compared to the original version of VQE, co-VQE has an advantage that its optimization process is by definition free of statistical or systematic error inherent in quantum hardware.

co-VQE does not exclude opportunities to benefit from quantum devices; one needs to rely on them to measure global quantities such as nonlocal order parameter and fidelity after the optimization.
As proof-of-concepts for our method, we present numerical simulations on 1D and 2D quantum spin models with topological phases.
First, we solve the 1D cluster model by co-VQE. 
We detect the topological phase transition by evaluating the nonlocal order parameter.
In addition, we demonstrate that even without prior knowledge of the order parameter, we could also identify the phases by applying clustering technique to fidelity.
Then, given that the ground states and nonlocal observables in 1D models are often achievable by classical computers, we also study the 2D toric code model with co-VQE.
We find that the ansatz initiated from the toric code state works well.
The important thing is that it may cover the region within the topological phase with a constant depth.
This leads to an expectation that we can derive advantages from quantum computers if we conduct co-VQE using a real quantum device for larger system sizes.

We expect that because of its error-immunity in the optimization process, co-VQE may have more potential than the original VQE to leverage NISQ devices to solve quantum many-body problems.
Our study is also suggestive of an intriguing field of quantum machine learning that combines measurement on quantum computers with classical machine learning to analyze quantum many-body problems.

\section*{Acknowledgements}
K.M. is supported by JST PRESTO Grant No. JPMJPR2019 and JSPS KAKENHI Grant No. 20K22330.
K.F. is supported by JST ERATO Grant No. JPMJER1601 and JST CREST Grant No. JPMJCR1673.
This work is supported by MEXT Quantum Leap Flagship Program (MEXTQLEAP) Grant No. JPMXS0118067394 and JPMXS0120319794.
We also acknowledge support from JSTCOI-NEXT program.

\bibliography{citation_vqe_topo} 

\end{document}